\documentclass[11pt]{article}

\input psfig.sty

\titlepage
\begin{document}

\vskip 3.5cm

\centerline{\bf \large \large \large Scientific Productivity with X-ray All-Sky Monitors}

\vskip 2.5cm

\centerline{\large \large Ronald A. Remillard, Alan M. Levine,}
\vskip 0.5cm
\centerline{\large \it Kavli Institute for Astrophysics \& Space Research, Massachusetts Institute of Technology}
\vskip 2.5cm
\centerline{\large \large \& Jeffrey E. McClintock}
\vskip 0.5cm
\centerline{\large \it Harvard-Smithsonian Center for Astrophysics}

\vskip 2.5cm

\centerline{\large Science white paper prepared for}
\vskip 0.5cm
\centerline{\large The Astronomy \& Astrophysics Decadal Survey Committee}
\vskip 0.5cm
\centerline{\large Panel on Stars \& Stellar Evolution}

\newpage

\section{Bright X-ray Sources and their Variability}

We address the value of X-ray monitoring with a summary of the lessons
learned from the {\it RXTE} All-Sky Monitor (ASM). We then describe
goals and opportunities for progress in the next decade.  The primary
functions of X-ray monitors are (1) to find transients, so that they
can be observed in detail, (2) to make light curves in different
energy bands to support a variety of analysis tasks for the different
classes of sources, and (3) to provide context for brief and
specialized observations by other instruments.

The brightest celestial X-ray sources include many examples of
accreting black holes (BHs), both stellar scale and supermassive
(SMBH), and accreting neutron stars (NSs) in a variety of evolutionary
conditions. These sources are used to address detailed questions
concerning BH physical properties, the structure of NSs, the origin of
relativistic jets in both binary systems and active galactic nuclei
(AGN), specialized areas of cosmology (e.g., the warm-hot
intergalactic medium), and the behavior of matter subjected to extreme
temperatures, intense magnetic fields, or strong gravity.  These
topics are central to NASA's historical contributions to astrophysics,
and they have remained cornerstones for science progress, e.g., in the
priorities expressed in the Roadmap {\it ``Beyond Einstein''}.

The X-ray sky is replete with variability. Many forefront themes in
astrophysics depend on the response to windows of opportunity defined
by unpredictable changes in flux or spectral characteristics.  For
example, it is believed that there are $\sim 10^8$
stellar-mass BHs in the Galaxy.  However, our knowledge about these
objects depends upon a very small number of mass-exchange binaries.
X-ray eruptions lead to their discovery, and ensuing dynamical
measurements of the companion star may indicate a compact-object that
is too massive to be a NS.  To date there are 18 known BHs in the
Milky Way, plus 4 others in local galaxies (Remillard \& McClintock
2006; Orosz et al. 2007; Prestwich et al. 2007).  There are an
additional $\sim$25 BH ``candidates'' in the Milky Way, based on the
similarity of X-ray timing and spectral properties.  Remarkably, 17 of
the BH systems in the Milky Way and 21 of the BH candidates are X-ray
transients (often recurrent).  Random instrument pointings would
usually find them in a quiescent state, which is a factor of $10^6$
below the outburst maximum (McClintock \& Remillard 2006). And even
within a given outburst, there are state transitions associated with
steady or impulsive types of relativistic jets (Fender 2006).

Many other types of X-ray sources are known to vary by a factor of 10
or more.  Transients account for the majority of known NS accretors
with low magnetic fields (so-called atoll and Z sources), as well as
those with strong fields (i.e., classical X-ray pulsars).
Furthermore, all 10 of the known X-ray millisecond pulsars are X-ray
transients. These sources are an evolutionary bridge between accreting
NS and millisecond radio pulsars, and the knowledge of their spin
period is crucial for interpretations of X-ray burst oscillations
(Watts et al. 2008), and for investigations of kHz quasi-periodic
oscillations (e.g., Wijnands 2006).

Opportunities are also derived from X-ray variations in classes quite
different from X-ray binaries. In the ``blazar'' (jet-dominated)
type of AGN, major flares prompt multi-frequency observations that
investigate the new ejections with coverage from radio to TeV gamma
rays. We can further extend this discussion to include topics
such as gamma ray bursts, tidal disruption events (i.e. stellar
infall) into SMBHs in non-active galaxies, crustal-quake activity in
magnetars, flares from stars with active coronae, and the poorly
understood group of fast X-ray transients.

\section{Statistics and Examples of X-ray Transients}

The ASM archive provides light curves (1996-2009) for
persistent sources and many transients discovered with {\it RXTE,
BeppoSAX, INTEGRAL, and Swift}. For discussions of statistics, we
adopt an X-ray flux threshold of 10 mCrab (or $2.4 \times 10^{-10}$
erg cm$^{-2}$ s$^{-1}$ at 2-10 keV), which is about one thousand times
fainter than the brightest source, Scorpius X-1. On average, our
threshold is exceeded by $\sim$20 outbursts per year from 12 different
sources, and $\sim$4 of these are seen as new discoveries for
astronomy. Fig.~1 illustrates some of the activity of the X-ray sky
during a roughly two year interval chosen at the center of the present
decade.  Each panel in Fig.~1 shows transients for different types
of accreting binary systems.

\begin{figure*}[htb]
\hspace*{0.25cm} 
\psfig{file=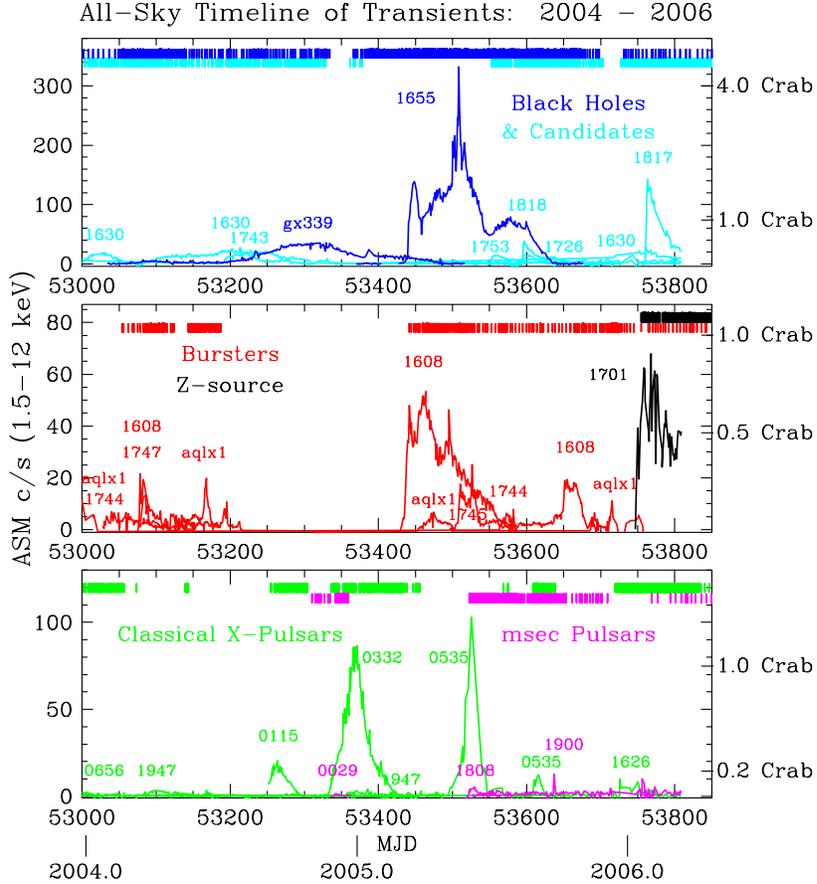,width=12.5cm}
\caption{ASM light curves for selected types of transients,
sampling 2004-2006.  The top panel illustrates dynamical BHs (blue)
and candidates (cyan). The middle panel shows NS binaries of atoll
type (red) and the unique Z-type transient (black).  The lower panel
shows classical accretion-powered pulsars (green) and X-ray
millisecond pulsars (magenta). Vertical ticks at the top of each
panel show the times of {\it RXTE} pointed observations.
\label{fig:asmlc}}
\end{figure*}

Some noteworthy examples in our slice of the transients time-line are
described as follows.  The extremely bright recurrence of the
dynamical BH binary GRO~J1655-40 (Fig.~1, top panel, dark blue)
provided new detections for the pair of high-frequency QPOs at 300 and
450 Hz (Homan et al. 2005), providing a strong message that these
frequencies (first displayed in 1996) are an inherent signature of the
accreting BH in that system.  This source is one of the BHs for which
a spin measurement ($a_* \sim 0.75$) has been derived from X-ray
continuum analyses of observations in the thermal state (Shafee et
al. 2006).  In addition, observations with {\it Chandra}
high-resolution gratings caught this source in a state where
absorption features from an ionized wind were discovered and
attributed to a strongly magnetized accretion disk (Miller et
al. 2006).

The discovery of XTE~J1701-462 (Fig.~1, middle panel, black) marked
the first-ever appearance of a transient with the timing properties of
a Z-type source (Homan et al. 2008).  The large dynamic range in the
luminosity of this X-ray transient provided a dramatic demonstration
that the behavioral sequence -- Cyg-like Z-source to Sco-like Z-source
to atoll source -- is a simple consequence of a decreasing mass
accretion rate. Detailed spectral analyses of 866 pointed {\it RXTE}
observations of XTE~J1701-462 provided physical interpretations for
each of the 3 branches of Z sources (Lin et al. 2009). The conclusions
are now being evaluated for the persistent Z sources, and further
tests are being conducted with data from other instruments.

Finally, the X-ray pulsar V0332+53 (bottom panel, green, maximum near
2005.0) is only the second case to show 3 or more cyclotron lines in
the X-ray spectrum (Pottschmidt et al. 2005). These lines determine
the strength of the polar magnetic field to be $2.7 \times 10^{12}$
G. Followup observations with {\it INTEGRAL} showed line shifts (to
lower energy) that were linearly correlated with luminosity, allowing
interpretation in terms of the height of a radiation-dominated shock
above the NS surface (Tsygankov et al. 2006).

At a given moment, there are an average of 5 active transients that
are brighter than 10 mCrab. At this level, the X-ray sky is still
dominated by 66 persistent sources: 1 BH and 2 candidates in the Milky
Way, 2 BHs in the LMC, 34 atoll and Z sources, 16 X-ray pulsars, 3
supernova remnants, 1 diffuse galactic component, 3 AGN, and 4
clusters of galaxies. Accreting NS and BH systems thus constitute the
majority of the brightest X-ray sources.  The overall science themes
are interwoven between persistent sources and transients, and we note
the following results.  The catalogs of accreting BHs, weakly
magnetized NSs, and X-ray pulsars are now dominated by accumulated
transients, rather than persistent sources, as noted in \S1 (see Liu
et al. 2006; 2007).  Even among persistent X-ray binaries, flux
variations by factors in the range 2 to 10 are common, and the
observations at different luminosity are needed to test physical
models of accretion in each type of binary and X-ray state (van der
Klis 2006).

Some ground-based observatories also rely critically on X-ray light
curves. The TeV observatories (VERITAS, HESS, Cangaroo, etc.) require
long exposures and dark skies to detect microquasars and blazars, and
they have successfully used the ASM data to choose targets in outburst
(e.g., Krawczynski et al. 2004).

We conclude that the need for an all-sky monitor for
bright X-ray sources is based on the following cornerstones: the value
of transients, the importance of luminosity variations in X-ray
sources, and the need to apply specialized instrumental capabilities
to the science of accreting compact objects.

%
%
\section{Goals for an Advanced Monitor, 2010-2020}

Extensive use of ASM light curves and alert services in the past 13
years has clearly contributed to productivity in
astrophysics. However, we have not fully absorbed these lessons. There
is no mention of the critical role of wide-angle X-ray cameras in
previous NASA roadmaps, and we are in serious danger of losing the
capability to monitor the X-ray sky during the next decade.  It would
be an negligent to fly a mission such as the International X-ray
Observatory without ensuring X-ray monitoring services in the same
time frame.

In 2009, X-ray monitoring functions are still being carried out by the
ASM (2-12 keV).  There is also coverage in hard X-rays (10-100 keV)
provided by INTEGRAL (ESA) and the Swift BAT (15-150 keV). All of these
are ``coded-aperture'' cameras, and this remains the best imaging
technique for wide-angle coverage with a design that is simple and
cost-effective.

Currently, no future NASA missions will deploy wide-angle X-ray
cameras, but two international missions will do so.  The Japanese MAXI
experiment is slotted for the Japanese Experiment Module on the ISS
(2010?). It will sweep the sky once every 90 min with fan beams,
yielding arcmin positions and sensitivity to 2 mCrab in 1 day.  The
Indian ASTROSAT Mission (2010?)  includes a monitor designed after
the {\it RXTE} ASM, providing 3 arcmin positions and sensitivity to
10-50 mCrab in 1 day for known sources.

We urge the NASA Decadal Survey Committee on Stars and Stellar
Evolution to acknowledge the need for wide-angle X-ray cameras: (1) to
provide alert and monitoring services for astrophysics programs
seeking measurements of BHs, NS, and sources of jets, and (2) to
gather primary spectral and timing data for a wide range of X-ray
variations that enable detailed investigations in relativistic
astrophysics.

Performance goals for wide-angle X-ray cameras and the primary rationale
are as follows:
\begin{itemize}
\item 1 arcmin positions for new sources to facilitate identification and
multi-frequency observations.

\item minimum bandwidth of 2-15 keV to distinguish
thermal and nonthermal spectral components.

\item sensitivity to 1 mCrab ($3 \sigma$; known sources) per day, to
track transients into faint states and to monitor the brightest
extragalactic sources.

\item public archive for light curves and raw events (with sub-ms 
time resolution) to support diverse data analysis activities.
\end{itemize}

The avenues for improvement for an advanced instrument are simply the
quantity and quality of the data products. One key parameter is the
all-sky, average duty cycle ($\tau$) that can be achieved in the
instrument design.  The ASM observes with $\tau \sim 0.02$, while the
Wide Field Camera on {\it BeppoSAX} (Dutch-Italian; 1996-2002)
operated with $\tau \sim 0.01$. The {\it Swift} Burst Alert Telescope
(BAT), designed for gamma ray bursts (15-150 keV), now monitors the
sky with $\tau \sim 0.10$.  Such low coverage is insufficient to
capture critical, infrequent events of exceptional
interest. Additionally, the BAT responds only above 15 keV and is
therefore incapable of observing the spectra of BH and NS accretion
disks, and also NS boundary layer, which is crucial for understanding
accretion physics.

Values of $\tau$ for selected instruments are shown versus photon
energy in Fig.~2.  Also shown are the duty cycles for special {\it
RXTE} pointing campaigns on a single target.  The most intensive 
such effort reached $\tau = 0.19$ for the 2002 outburst of the
X-ray millisecond pulsar, SAX~J1808-3658.

{\bf \it {Argos-X} -- concept for an advanced all-sky X-ray
observatory:} \\
Fig.~2 also shows the design goal ($\tau = 0.5$) for
{\it Argos-X}, proposed as a NASA SMEX in 2008.
{\it Argos-X} would be the first observatory in any waveband to
instantaneously view half the sky, while providing arcmin positions.
The design consists of 25 cameras, each with field of view (FOV) of
$40^\circ \times 40^\circ$, FWHM.  The composite FOV covers the whole
sky, except for a $60^\circ$ circle intended as a solar exclusion
zone. A simple operations plan calls for stationary pointing during
observations, an equatorial orbit, and daily slews of $1^\circ$ to
maintain the solar axis. The average daily exposure would be 57 ks per
source, reaching 1 mCrab sensitivity ($4 \sigma$) for known sources.

\begin{figure*}[htb]
\hspace*{0.5cm} 
\psfig{file=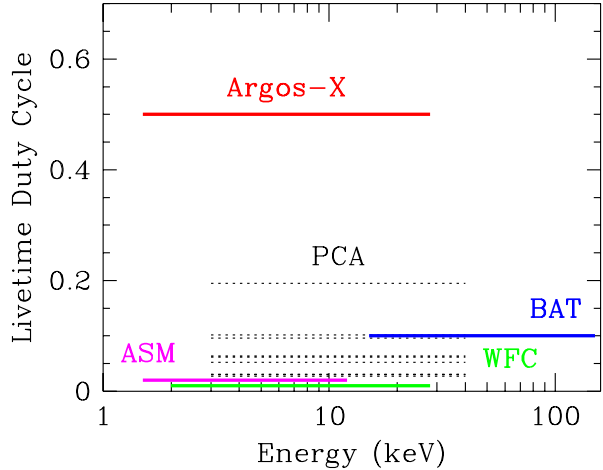,width=10cm}
\caption{Average all-sky live-time fraction for wide-angle
monitors (solid lines) or for the duty cycles of {\it RXTE} PCA pointings 
for noteworthy campaigns on a single source (dotted lines).
\label{fig:fig_merit}}
\end{figure*}

Achieving a high value of $\tau$ would provide rapid advances in our
understanding of cosmic explosions that occur on timescales of minutes
to an hour, e.g. fast X-ray novae and GRB-related X-ray flashes (Heise
\& in~'t Zand 2006).  Such sky coverage also creates extraordinary
synergy with other instruments: e.g. as the ``eyes'' for Advanced
LIGO, as it ``listens'' for gravitational waves.  There is also
opportunity for partnership with wide-angle radio observatories under
construction in each hemisphere: LOFAR (Netherlands) and the Murchison
Widefield Array (US-Australia). The joint observations would provide a
basis for solving the mysteries of relativistic jets (Fender 2006),
where the primary obstacle has been the inability to gather data at
those times when impulsive jets are launched.  Productive partnerships
would also be engaged with wide-angle optical facilities under
construction: the Large Synoptic Survey Telescope (LSST) and the
Panoramic Survey Telescope And Rapid Response System (Pan-STARRS).

The quality of primary data products from wide-angle X-ray cameras
further depends on the choice of the detector. Technology plays a
large role in defining options. Si pixel detectors are already
available with large pixel size (e.g. 2.5 mm), fast time-tagging
capability, broad sensitivity range (e.g. 1-30 keV for 0.5 mm pixel
depth), and good energy resolution (e.g. 600 eV with current designs
for low-power ASICs).  Further improvements are expected in the next
few years.  

Compared to the ASM, an advanced X-ray monitor could achieve tenfold
improvements in sensitivity and position resolution, while observing
with a broader energy range and $\tau$ increased by a factor of 25.
Si detectors can produce data products that rival the quality of
current pointed instruments.  For example, for bright sources it would
be possible to do the coded-mask spatial deconvolution every 3 ks,
with 128 spectral channels covering 1.5-30 keV, thereby providing
outstanding capabilities, e.g., for observing relativistic Fe lines in
BH sources. The primary data and followup observations would address
the following science goals:

\begin{itemize}

\item Find BHs in the Galaxy and conduct spectral analyses
of the continuum and Fe line to constrain mass and spin.

\item Capture state transitions in X-ray binaries to determine 
the nature of non-thermal X-ray states.

\item Capture accretion disk changes that cause impulsive relativistic jets in
Galactic microquasars.

\item Measure Fe line variations with luminosity or state to help
prepare for the International X-ray Observatory.

\item Measure the locations, spin, magnetic fields, and radiation properties
of NS in the Milky Way.

\item Use routine exposure times of $10^7$ s to determine binary periods and 
to study X-ray bursts and superbursts.

\item Survey the local universe for SMBHs that are quiescent (via
stellar infall events) or obscured from view (via detections at 5--20
keV).

\item Measure break frequencies in the power spectra of active galaxies and
relate these to the masses of their SMBHs.

\item Measure new ejections in the jets from ``blazar'' type AGN.

\item Provide time-critical information to other space missions and
ground-based observatories in order to increase their productivity.

\item Survey gamma ray bursts in the X-ray band and use them to
support efforts to detect gravitational waves with Advanced LIGO.

\end{itemize}

\section{References}

\small
\noindent Fender, R. 2006, in ``Compact Stellar X-ray Sources'', eds. \\
\hspace*{1.0cm} W. Lewin \& M. van der Klis, Cambridge U. Press, 381-420 \\
\noindent Heise, J. \& in~'t Zand, J. 2006, ibid., 267-278 \\
\noindent Homan, J., et al. 2005, AAS, 207, 102.01 \\
\noindent Homan, J., et al. 2007, ApJ, 656, 420 \\
\noindent Krawczynski, H., et al. 2004, ApJ, 601, 151 \\
\noindent Lin, D., Remillard, R.A., \& Homan, J. 2009, ApJ, in press;
arXiv:0901.0031 \\
\noindent Liu, Q.Z., van Paradijs, J., \& van den Heuvel, E.P.J. 2006, A\&A, 455, 1165 \\
\noindent Liu, Q.Z., van Paradijs, J., \& van den Heuvel, E.P.J. 2007, A\&A, 469, 807 \\
\noindent McClintock, J.~E., \& Remillard, R.~A. 2006, in ``Compact Stellar X-ray \\
\hspace*{1.0cm} Sources'', eds. W. Lewin \& M. van der Klis, Cambridge U. Press, 157-214 \\
\noindent Miller, J.M., et al. 2006, Nature, 441, 953 \\
\noindent Orosz, J.A., et al. 2007, Nature, 449, 872 \\
\noindent Pottschmidt, K., et al. 2005, ApJ, 634, L97 \\
\noindent Prestwich, et al. 2007, ApJ, 669, 21 \\
\noindent Remillard, R.A., \& McClintock, J.E. 2006, ARAA, 44, 49--92 \\
\noindent Shafee, R., McClintock, J.E., Narayan, R., Davis, S., Li, L.-X., \& Remillard, \\
\hspace*{1.0cm} R. 2006, Ap.J. 636, L113 \\
\noindent Tsygankov, S. S., Lutovinov, A. A., Churazov, E. M., \& Sunyaev, R. A. \\
\hspace*{1.0cm} 2006, MNRAS, 371, 19 \\
\noindent van der Klis, M. 2006, in ``Compact Stellar X-ray Sources'', eds. \\
\hspace*{1.0cm} W. Lewin \& M. van der Klis, Cambridge University Press, 39-112 \\
\noindent Watts, A.L., Patruno, A., \& van der Klis, M. 2008, ApJ, 688, L37 \\
\noindent Wijnands, R. 2006, AdSpR, 38, 2684 \\

\end{document}